\documentclass[aps,twocolumn,twoside,groupedaddress,dvips,floatfix]{revtex4}
\usepackage[]{graphicx}
\usepackage[latin1]{inputenc}
\usepackage[USenglish]{babel}
\usepackage[squaren]{SIunits}
\usepackage{mathptmx}
\usepackage{hyperref}

\begin{document}
\newcommand{\nm}{\nano\meter}
\newcommand{\dose}{\centi\meter^{-2}}
\newcommand{\keV}{\kilo\electronvolt}
\newcommand{\eV}{\electronvolt}

\renewcommand{\arraystretch}{1.3}

\hyphenation{nano-clusters}

\bibliographystyle{apsrev}


\title{Size and Location Control of Si Nanocrystals at Ion Beam Synthesis in Thin SiO$_2$ Films}


\author{Torsten Müller}
\email[]{T.Mueller@fz-rossendorf.de}
\thanks{Tel. +49 351 260 3148; Fax. +49 351 260 3285}
\homepage[]{http://www.fz-rossendorf.de}
\author{Karl-Heinz Heinig}
\author{Wolfhard Möller}
\affiliation{Forschungszentrum Rossendorf, Institut f\"{u}r Ionenstrahlphysik \\und 
Materialforschung, PO Box 51 01 19, 01314 Dresden, Germany}


\date{\today}

\begin{abstract}
Binary collision simulations of high-fluence \unit{1}{\keV} Si$^+$ ion implantation into \unit{8}{\nm} thick SiO$_2$ films on (001)Si were combined with kinetic Monte Carlo simulations of Si nanocrystal (NC) formation by phase separation during annealing. For nonvolatile memory applications, these simulations help to control size and location of NCs.
For low concentrations of implanted Si, NCs form via nucleation, growth and Ostwald ripening, whereas for high concentrations Si separates by spinodal decomposition. 
In both regimes, NCs form above a thin NC free oxide layer at the SiO$_2$/Si interface. This, self-adjusted layer has just a thickness appropriate for NC charging by direct electron tunneling. Only in the nucleation regime the width of the tunneling oxide and the mean NC diameter remain constant during a long annealing period. This behavior originates from the competition of Ostwald ripening and Si loss to the Si/SiO$_2$ interface. The process simulations predict that, for nonvolatile memories, the technological demands on NC synthesis are fulfilled best in the nucleation regime.
\end{abstract}

\pacs{}
\keywords{nanocrystals, phase separation, nonvolatile memory, nucleation, spinodal decomposition}

\maketitle

Recently, nonvolatile memory concepts based on nanocrystals (NCs) embedded in the gate oxide of MOS transistors have attracted much interest \cite{Tiwari:1996}. For that aim, NCs have been synthesized by a variety of techniques like chemical vapor deposition \cite{Takahashi:2000}, ion implantation \cite{Normand:2001,Borany:1999}, and Si aerosol deposition \cite{Ostraat:2001}. Ion implantation followed by thermally activated precipitation of the implanted impurity atoms is most compatible with current silicon technology. By low-energy Si$^+$ ion implantation into thin SiO$_2$ layers on (001)Si, NCs of Si were formed a few nanometers above the Si/SiO$_2$ interface \cite{Normand:2001}. This allows charging of the NCs by direct electron tunneling, which is a prerequisite for high endurance and low operation voltages \cite{DeSalvo:2001}. 
\enlargethispage*{2em}
Further optimization of location and size of ion beam synthesized NCs for memory application requires a deeper understanding of the mechanisms involved, which determine (i) the built-up of Si supersaturation by high-fluence ion implantation and (ii) NC formation by phase separation. For that aim, process simulations were divided into two steps. The Si implantation was studied using the binary collision code TRIDYN \cite{Moeller:1984}, which includes dynamic target changes. The phase separation of Si from SiO$_2$ during subsequent annealing has been simulated with a kinetic lattice Monte-Carlo code, which describes the thermally activated processes. 
\begin{figure}[t]
 \includegraphics[width=86mm]{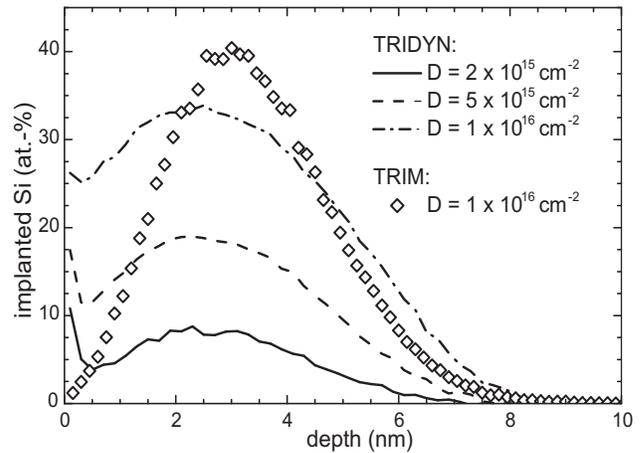}
 \caption{TRIDYN depth profiles for \unit{1}{\keV} Si$^+$ implantation into SiO$_2$. For comparison a TRIM profile has been added.}
 \label{fig:tridyn1}
\end{figure}   

The TRIDYN depth profiles are shown in Fig.~\ref{fig:tridyn1} for \unit{1}{\keV} Si$^+$ ion implantation into SiO$_2$. TRIDYN takes into account dynamic target changes due to ion deposition, ion erosion and ion beam mixing. The input parameters required by the simulation include the displacement and surface binding energies of target atoms. The displacement energies $E_d$ for both, Si and O, were assumed to be \unit{8}{\eV}. 
This value proved to yield satisfactory agreement between earlier TRIDYN simulations of ion mixing and experiments \cite{Moeller:1986} and is also consistent with the choice of Sigmund and Gras-Marti \cite{Sigmund:1981} in their theoretical treatment of ion mixing. For the present problem, simulations with $E_d$ varied by factors of 0.5 and 2 did not show any significant differences of the Si deposition profiles.
The surface binding energies of Si and O are assumed to vary linearly with the surface composition in a way that they balance the enthalpies of sublimation and decomposition of Si and SiO$_2$, respectively \cite{TRIDYN:2001,Kelly:1980}.

The Si profile broadens by ion beam mixing, sputtering and swelling. For comparison, a Si implantation profile calculated by TRIM \cite{ZBL:1985} has been added to Fig.~\ref{fig:tridyn1}, which is much sharper than the corresponding TRIDYN profile. Accordingly, the Si peak concentration is significantly overestimated by TRIM. Additionally, sputtering has led to a Si enrichment at the target surface.

\begin{figure}[tb]
 \includegraphics[width=86mm]{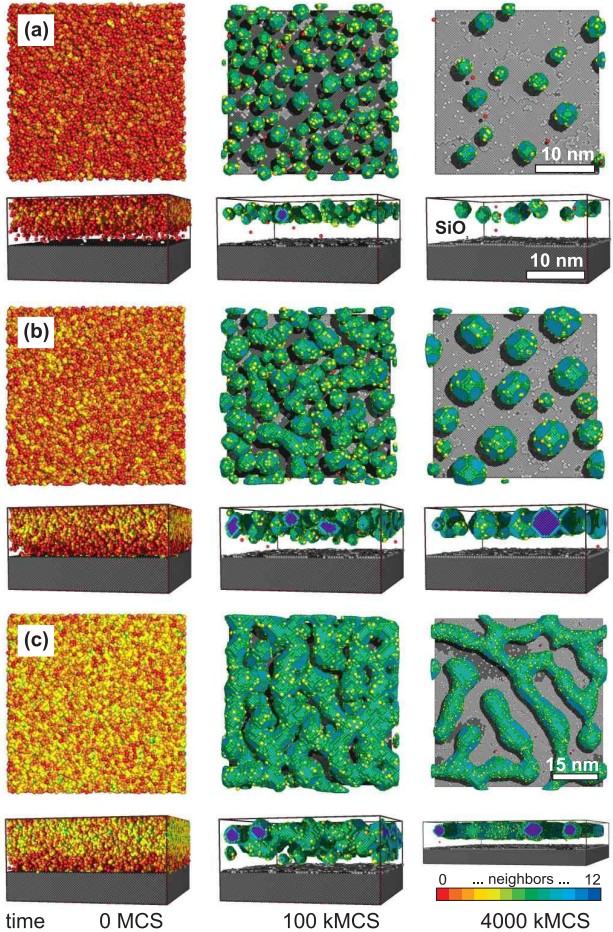}
 \caption{Snapshots of KMC simulations (top view and cross-section) of phase separation in \unit{8}{\nm} thick SiO$_2$ on (001) Si during annealing. 
The simulations start from \unit{1}{\keV} Si$^+$ TRIDYN profiles for fluences of (a)  \unit{2 \times 10^{15}}{\dose}, (b) \unit{5 \times 10^{15}}{\dose}, and (c) \unit{1 \times 10^{16}}{\dose}, respectively. Two regimes were identified, "nucleation and growth" (a) and "spinodal decomposition" (b),(c). Additionally, percolation is observed at the highest fluence (c). Atoms are colored according to their coordination. The \unit{15}{\nm} scale only applies for the lower right corner.}
 \label{fig:KMC}
\end{figure}

During subsequent annealing, Si implanted into the thin SiO$_2$ separates from the oxide phase. In general, the process of phase separation is expected to be a sequence of physical mechanisms like nucleation, growth, and Ostwald ripening of Si precipitates or, at higher Si concentrations, spinodal decomposition and interfacial energy minimization of the Si/SiO$_2$ mixture, respectively. However, these mechanisms are the result of a variety of elementary events (like bond breaking, diffusional jumps of atoms, chemical reactions etc.) that occur in random sequence. Here, they are studied by a kinetic 3D lattice Monte Carlo (KMC) method, which is discussed elsewhere in detail \cite{Strobel:2001}.

The kinetics of Si atoms is described in a solid host matrix (SiO$_2$), which is the background or "system's vacuum". Thereby, an underlying fcc lattice has been assumed, which is the most isotropic lattice. Within this host dissolved Si diffuses and can form precipitates. (The lattice spacing was chosen such that the correct atomic Si density is obtained.)  
Applying the classical lattice gas model with attractive Si-Si interaction, the energetics is determined by the nearest-neighbor Ising model. The Metropolis algorithm \cite{Metropolis:1953} is used to describe the kinetics of the system.

Si dissolved in the matrix performs diffusional jumps from one lattice site to another with the probability $1/\tau \exp \left\{ - E_A/k_B T \right\}$, where $E_A$ is the activation energy of diffusion, $1/\tau$ is the attempt frequency and $k_B T$ has its usual meaning. Due to a time scale normalization, the internal time unit of the KMC simulation is a Monte Carlo step (MCS) given by $\tau \exp\{E_A/k_B T\}$, thus depending on the temperature. Si dimers, trimers and larger agglomerates can form and exhibit a binding energy, which is the product of the Si-Si bond strength $E_B$ by the number of Si-Si bonds. A Si atom having $n_i$ Si neighbors jumps to an empty neighboring site having $n_f$ Si neighbors with a reduced probability $1/\tau \exp \left\{ - \left[E_A+(n_i-n_f)E_B\right]/k_B T \right\}$ if $n_f > n_i$. Otherwise, the diffusional jump probability remains valid.

In principle, the bond strength $E_B$ for our KMC simulation can be determined from the solubility of Si in SiO$_2$ via the detailed balance of Si attachment/detachment at the Si/SiO$_2$ interface. (It has to be taken into account that the coordination number in the fcc lattice is 12 instead of 4 in the case of Si lattice.) However, the diffusivity and the solubility of Si in SiO$_2$ are largely unknown. Thus, a direct relation of simulation and experimental temperatures is difficult, which holds for the time scale too. Nevertheless, the path of systems's evolution towards equilibrium and the regimes of phase separation predicted by KMC simulations may improve the process understanding substantially.

\begin{figure}[tb]
 \includegraphics[width=86mm]{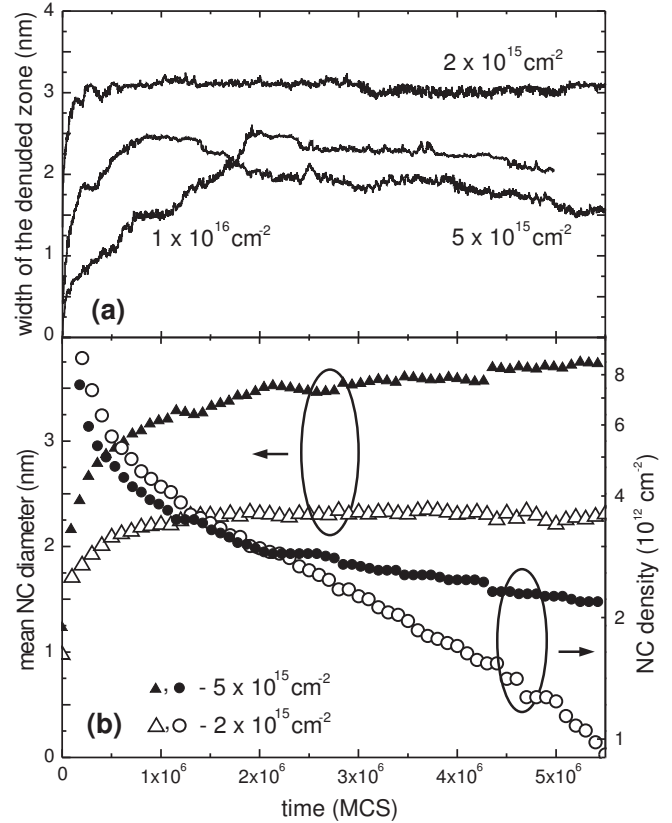}
 \caption{Evolution of the width of the denuded zone (a) as well as the mean NC diameter and density (b) during annealing for both regimes, nucleation and spinodal decomposition.} 
 \label{fig:size}
\end{figure}

In Fig.~\ref{fig:KMC}, snapshots of KMC simulations of phase separation are shown for $E_B/k_B T = 2.0$. Thereby, the TRIDYN profiles of Fig.~\ref{fig:tridyn1} were used as initial Si distributions. The size of the simulation cell is \unit{56 \times 56 \times \Delta z}{\nm^3}, where $\Delta z$ is \unit{8}{\nm} plus swelling due to implantation. Besides the image at the lower right corner, a quarter of the simulation cell is shown only. The simulation cell borders on a fixed (001) layer of the Si substrate to account for the Si/SiO$_2$ interface. Periodic boundary conditions were applied in the oxide plane, whereas reflecting boundary conditions were assumed at the surface. 

Two regimes of phase separation are predicted by the KMC simulations. A "nucleation and growth" regime is observed for \unit{2 \times 10^{15}}{Si^+ \dose}. Si NCs form by nucleation and grow further at the expense of Si supersaturation (Fig.~\ref{fig:KMC}~(a), 100 kMCS). 
Later on, NCs grow by Ostwald ripening and finally dissolve by Si loss to the SiO$_2$/Si interface.
Above \unit{2 \times 10^{15}}{Si^+ \dose} a "spinodal decomposition" regime is identified, where due to the vanishing nucleation barrier \cite{LeGoues:1984} non-spherical, elongated Si structures are formed (Fig.~\ref{fig:KMC}~(b), 100 kMCS). At even higher Si concentrations (\unit{1 \times 10^{16}}{\dose}), above the percolation threshold, the phase separated Si becomes laterally connected (Fig.~\ref{fig:KMC}~(c), 100 kMCS). This network of Si does not decay into droplets during longer annealing (Fig.~\ref{fig:KMC}~(c), 4000 kMCS). An electrical charge brought to this network can spread over several tens of nanometer, i.e.~the phase separated Si behaves like a floating gate in a conventional metal-oxide-silicon transistor.

Below the percolation threshold at \unit{5 \times 10^{15}}{\dose}, the initially non-spherical Si structures evolve into spherical NCs, which can hardly be distinguished at this late stage from that formed by nucleation and growth (see Fig.~\ref{fig:KMC} (a),(b), 4000 kMCS). 
The SiO$_2$/Si interface acts as an effective sink for Si in both regimes, which results in a zone denuded of NCs. However, a more detailed consideration reveals some differences between the nucleation regime and the spinodal decomposition regime. As shown in Fig.~\ref{fig:size} a), for the nucleation regime (\unit{2 \times 10^{15}}{\dose}) the width of the denuded zone is constant over long annealing times. A similar dependence is observed for the NC size (Fig.~\ref{fig:size} (b)). The competition between Ostwald ripening and Si loss to the Si/SiO$_2$ interface keeps the mean NC diameter constant over a long period of annealing. The Si loss manifests itself in a rapid decrease in NC density (Fig.~\ref{fig:size} (b)).
On the other hand, for the spinodal decomposition regime(\unit{\geq 5 \times 10^{15}}{\dose}), the interface minimization of the non-spherical Si structures leads to a narrowing of the denuded zone (Fig.~\ref{fig:size} (a)). Moreover, Ostwald ripening is more effective than Si loss to the interface and, hence, up to \unit{4000}{kMCS} the mean NC size is increasing as can be seen in Fig.~\ref{fig:size} (b). In contrast to the nucleation regime, the NCs become larger and dissolve slower ( Fig.~\ref{fig:size} (b)). Of course, a long lasting annealing results also in this regime in a complete dissolution of NCs.

It should be emphasized that NCs form behind a zone completely denuded of NCs (see cross section views of Fig,~\ref{fig:KMC}).  Additionally, the distance of the NCs from the interface is small enough to allow their charging by direct electron tunneling. This self-alignment of NCs as well as their degradation free charging/decharging is crucial for application in nonvolatile memories \cite{DeSalvo:2001}. 
Varying ion implantation energy and annealing temperature gives additional control over the width of the depleted zone and the NC size, which will be described elsewhere \cite{Mueller:2002}. Using Si implantation profiles predicted by TRIM instead of TRIDYN lead to a strong overestimation of the percolation behavior during phase separation, which will be shown there too. 

In conclusion, two regimes of Si NC formation by 1 keV Si$^+$ ion implantation into thin SiO$_2$ and subsequent annealing have been found by process simulation. Below a Si fluence of \unit{2 \times 10^{15}}{\dose}, NCs form by nucleation and growth, while at higher fluences spinodal decomposition occurs. At \unit{1 \times 10^{16}}{\dose}, percolation leads to a spatially connected 2D pattern of Si in SiO$_2$. 
The present KMC simulations predict that NC fabrication for nonvolatile memory applications should be performed preferably in the nucleation regime. In this regime, the width of the denuded zone does not depend critically on the annealing time and/or temperature. The synthesized NCs are small (\unit{2 .. 3}{\nm} diameter) and of high area density ($> 10^{12}$ cm$^{-2}$). 

This work was sponsored by the European Community under the auspices of the GROWTH project GRD1-2000-25619. 

\bibliography{Phase-sep}

\newpage

\end{document}